\documentclass{ws-ijmpa}

\begin{document}

\markboth{F. Cianfrani, O. M. Lecian}
{E.C.G. Stueckelberg: a forerunner of modern physics II}

%
\catchline{}{}{}{}{}
%

\title{Stueckelberg: a forerunner of modern physics II
}

\author{FRANCESCO CIANFRANI}

\address{ICRA---International Center for Relativistic Astrophysics\\ 
Dipartimento di Fisica (G9),
Universit\`a  di Roma, ``Sapienza'',\\
Piazzale Aldo Moro 5, 00185 Rome, Italy.\\ 
francesco.cianfrani@icra.it}

\author{ORCHIDEA MARIA LECIAN}

\address{ICRA---International Center for Relativistic Astrophysics\\ 
Dipartimento di Fisica (G9),
Universit\`a  di Roma, ``Sapienza'',\\ 
Piazzale Aldo Moro 5, 00185 Rome, Italy.\\ 
lecian@icra.it}

\maketitle

\begin{history}
\received{Day Month Year}
\revised{Day Month Year}
\end{history}

\begin{abstract}

We will investigate some aspects of Stueckelberg's work, which have contributed to the development of modern physics. On the one hand, the definition of diffuse boundaries in the calculation of scattering amplitudes will be reviewed, and compared with the other proposals by physicists of that time. On the other hand, the applications of Stueckelberg's description of a massive vector field in the Standard Model will be discussed.


\end{abstract}

\ccode{PACS number: 12.60.Cn}

\section{Introduction}
Stueckelberg's activity in Theoretical Physics was characterized by brilliant intuitions, which made his results a basis for future developments. Among his achievements, it is worth noting the definition he proposed for ``in'' and ``out'' states and the development of a quantum theory for massive vector fields. 

\vspace{0.5cm}

Stueckelberg's definition of ''in'' and ''out'' states is a valuable attempt to encode the time uncertainties of the initial and final observation epochs without affecting the unitarity of the theory. The whole experimental system, consisting of microscopic particles and macroscopic counters, is considered. Within this framework, observations are described as taking place in a finite region within two layers, rather than within two surfaces. As a result, non-conservative processes are allowed but highly suppressed, and the limit of massless electrodynamics is recovered by pushing the boundaries towards infinity.\\
As far as vector fields are concerned, Stueckelberg pointed out that the Proca formulation is not satisfactory and some sort of gauge symmetry has to be maintained also for massive fields. This perspective is really surprising for his time, since the decisive role of symmetries in view of removing divergences by renormalization would be clear more than 20 years later. According with this point of view, a scalar field enters the mathematical description, and a proper gauge invariance reduces the number of degrees of freedom to three \cite{Stu,Stu2}. In 1972, Lowenstein and Schroer \cite{LB} probed that renormalizability and unitarity stand for the Stueckelberg formulation, which was thus recognized as an alternative of the spontaneous symmetry breaking mechanism in view of giving mass to Abelian fields.\\

In the second Section, the main feature of the diffuse-boundary formalism will be discussed, together with a comparison with the canonical Tomonaga-Schwinger radiation theory. In the third Section, after a brief review of the Stueckelberg mechanism describing massive vector fields, recent applications of this approach to the Standard Model will be outlined.

\section{Boundary states in quantum field theory}
We will revise Stueckelberg's approach to quantum field theory (QFT) and the definition of boundary states: his definition of ''in'' and ''out'' states has to be compared to and interpreted according to the speculations of contemporary physicists.

\subsection{The Tomonaga-Schwinger radiation theory}
In non-relativistic quantum electro-dynamics, the dynamical variables are the electromagnetic potential $A_{\mu}(\vec{r})$ and the spinor electron-positron field $\psi_{\alpha}(\vec{r})$, all behaving as time-independent operators on the state vector $\Phi$ of the system. Given any 3D space-like surface $\sigma=(t(\vec{r}), \vec{r})$, the observation of the system on $\sigma$, $\Phi(\sigma)$ is given by $i\partial\Phi(\sigma)/\partial \sigma(\vec{r})=H(\vec{r})\Phi(\sigma)$, where $H(\vec{r})$ is the operator which accounts for the energy density at the point $\vec{r}$. The results of measurements of field quantities at any given point of space are independent of time, for any constant state-vector $\Phi_{0}$.

In relativistic quantum mechanics, a new relativistic-invariant state-vector $\Psi$, which accounts for photons, electrons an positrons traveling freely through space without interaction or external disturbance, must be introduced. The total energy density $H(\vec{r})=H_{0}(\vec{r})+H_{1}(\vec{r})$ account for free fields $H_{0}(\vec{r})$ and for their interaction $H_{1}(\vec{r})$. The Tomonaga-Schwinger form of the Schroedinger equation \cite{t46} is the differential form of that of the previous case, which can be written as $\partial\Psi/\partial \sigma(x_{0})=H_{1}(x_{0})\Psi$: after the introduction of perturbation theory, it can be solved explicitly, bringing results identical to those of the Heisenberg $S$ matrix.
  
\subsection{Diffuse boundaries}
In Reff. \refcite{st1}\refcite{st2}, Stueckelberg deals with the elimination of arbitrary constants in relativistic QFT. In particular, he analyzes boundary divergences, due to the sharp space-time limitation of the space-time region $V$ where interactions take place. Boundary effects involve a classical field, which describes the action of the counters necessary to distinguish between incoming and outgoing particles. Particle creation, annihilation and collision are influenced by macroscopic fields (experimental devices). The action accounting for such interactions must must be specified in a 4D region $V$ (finite volume, and time period where macroscopic matter does not show up). For experimental purposes, $V$ must be contained in a 4D layer of classically-described macroscopic matter (counters), which records incoming and outgoing distributions. Such a macroscopic field interacts with elementary particles through processes where energy-momentum is modified.

As a Gedankenexperiment, S. proposes the situation of macroscopic fields that interact with elementary particles only during two very short periods, before and after the quantum interaction. Such counters can be achieved by varying at leisure the coupling constant of the interaction between elementary particles and the counters. The limiting process is described by such counter layers becoming hypersurfaces, i.e., the experimental realization of differential theory. Even if this limit existed, the effects of such time (un)certainty would provide an arbitrarily large energy available for the process.

As an alternative scheme, the notion of {\it ideal counter} can be introduced. Such a counter is able to specify exactly the number of incoming and outgoing particles, $M$ and $N$, respectively, in the region $V$. The operator $S[V]=S[V,N/M]$, such that
\begin{equation}
\psi''=S[V]\psi'
\end{equation} 
is a unitary matrix, with a functional dependence on the region $V$. The process takes place inside the space-time region $V$, separated from the exterior space, characterized by $V(x)=0$, by a layer $F$ , described by $\partial_{\alpha}F\neq0$. Only the case where the interaction region $V$ is much larger than the counter thickness $F$ is of physical relevance. In this case, it possible to distinguish between conservative processes (which take place inside $V$) and non-conservative processes ( which take place inside the layers $F$ or close to them, and where the energy uncertainty $\Delta E$ is involved in the energy-conservation balance). If the limiting process exists, the final state $\psi''$ can be considered as a functional of the final time $t''$. If the limit does not exist, the idea of the final state $\psi''$ at the time $t''$ should be abandoned, and the description of the process should be treated under an integral theory. Anyhow, the transition amplitudes evaluated for sharp boundaries would bring divergent results. These divergences can be eliminated by the introduction of {\it diffuse boundaries} \cite{st3}.

Within this framework, time does not appear as a parameter, $t''-t'=2T$, but the initial and final epochs have finite duration, $\Delta t'$ and $\Delta t''$. While, in the Tomonaga-Schwinger theory, the state vector is a functional of a hypersurface, within this formalism it is a functional of a layer $V(x)$, which appears in the integration formulas as $\int d^{4}xV(x)$, where $V$ is given in term of two layers $F(x)$, $V(x)=F'(x)-F''(x)$, so that the pertinent transitions read
\begin{equation}
\Psi[F''(x)]=S[V]\Psi[F'(x)],
\end{equation}
and the probability amplitude for the time measurement $f'_{\alpha}(x)$ reads $f'_{\alpha}(x)=\partial_{\alpha}F'(x)$.

The formalism of diffuse boundaries applies also to those cases ( {\it non-conservative processes}), for which Standard QED predicts a vanishing probability. As an example, the emission of a massive photon by a free electron can be analyzed. In this case, the probability amplitude for the time measurement reduces to a function of time only, i.e., $f'_{i}(x)=0$, $f'_{0}(x)\equiv f'(t)$, and its Fourier transform $g(\omega)$, normalized as $\mid g'(0)\mid=\mid g''(0)\mid=1$, describes the precision by which the time $t$ has been taken in $\Delta t$, and allows for the definition of the Fourier transform of the region $V$,
\begin{equation}
V(p)=i(2\pi\omega)^{-1}\delta(\vec{p})[g'(\omega)-g''(\omega)],
\end{equation} 
so that setting $V(p)\equiv\delta(p)$ is equivalent to extending the integration domain over an infinite space-time region.

The probability density that a massive photon of frequency $(\omega, \omega+d\omega)$ has been emitted by a free electron reads
\begin{equation}\label{omega}
d(\omega)=d\omega\left(\mid g'(\omega)\mid^{2}+\mid g''(\omega)\mid^{2}\right)n(\omega)=d\omega'(\omega)+d\omega''(\omega)
\end{equation}
and can be interpreted as the sum of two probabilities referred to as taking place only during the initial and final observation, for which the two probability amplitudes $g'(\omega)$ and $g''(\omega)$ play the role of convergence factors.

Since the process is non-conservative, the additional energy $\omega$ can be interpreted as being provided by the experimental device during either the initial or final period of observation, with probability $\mid g'(\omega)\mid^{2}$ and $\mid g''(\omega)\mid^{2}$, respectively. The choice $f(t)\equiv\delta(t)\Rightarrow g(\omega)\equiv\exp(i\omega t)$, corresponding to sharply-defined time boundaries, would bring a divergent result for (\ref{omega}).

If the period $2T=t''-t'$ between the initial and final epochs is long with respect to their uncertainties, $\Delta t'$ and $\Delta t''$, a frequency $\omega_{0}$ may be defined,
\begin{equation}
2T>>\omega_{0}^{-1}>>t'', t',
\end{equation}
such that the following substitutions are possible
\begin{equation}
\mid g''(\omega)-g'(\omega)\mid^{2}=\left(2Tsin\omega T\right)^{2}, \quad\omega^{2}<\omega_{0}^{2}
\end{equation}
\begin{equation}
\mid g''(\omega)-g'(\omega)\mid^{2}=\left( \mid g''(\omega)\mid^{2}+\mid g'(\omega)\mid^{2}\right) \quad\omega^{2}>\omega_{0}^{2}.
\end{equation}

In this case, the transition amplitude for this non-conservative transition splits up into two parts: a time-dependent term,which is affected by the presence of the photon mass $\mu$, and a time-independent term, which decomposes into two contributions, due to the two limits of the region $V(x)$.

Furthermore, it is easy to recognize that this process has a straightforward classical analogue, according to which the expectation value of the photon field correspond to a classical field, which accounts for the point-like static field of the electron at the time of observation $t\sim t'$ everywhere, but within an infinitesimal sphere, whose radius is of order of the uncertainty $\Delta t'$ of the time measurement.

The influence of the counters delimiting $V(x)$ can be made negligible if $V$ is much larger than the hyper-layer itself, i.e.,
\begin{equation}
2T>>\Delta t'', \quad\Delta t'>>\mu^{-1}:
\end{equation} 
this way, {\it non-conservative processes are allowed, but highly suppressed}. The limit $\mu\rightarrow 0$, i.e., the limit of massless QED, pushes the time boundaries of the observational epochs towards infinity.

\subsection{The intermediate representation}
The problem of an operationally-motivated description of initial and final states in QFT was one the main themes of the then scientific unrest. In fact, in the same year, an interesting paper by F. Dyson appeared \cite{d51}, in which a similar problem was analyzed.

The {\it intermediate representation} (between the interaction and the Heisenberg representations) is defined by constructing explicitly a unitary operator that transforms the state vector of the interaction representation into the state vector of the new representation. In this representation, changes of the state of a system are described in the low frequency by changes in the state vector, as in the interaction representation, and in the high frequency by the time variations of the field operators, as in the Heisenberg representation. Heisenberg operators can this way be regarded as special limiting cases of the intermediate representation. The averages of the Heisenberg operators in QED over finite space-time regions are divergence-free after renormalization.

Integrating over finite space-time regions is the starting point of both Dyson's and Stueckelberg's works, but, unfortunately, the two methods are so different that a direct comparison is hardly possible, as pointed out by Dyson himself \cite{d51}.  Nevertheless, it can be useful for our purposes to note how the definition of initial and final states as functional of a finite space-time region and its application in the elimination of divergences was an open problem at that time.

\subsection{Discussion}
A few years later, the idea of {\it asymptotic states} was proposed \cite{mah}. In this case, the initial state $\phi_{in}(x)$ is a limit of the interacting field $\phi(x)$, and the coupling terms in the equations of motion are described by an adiabatic cut-off equal to one at finite times, but vanishing as $\mid t\mid\rightarrow\infty$: physical quantities are expressed in the limit of the removal of the cut-off $Z$, i.e., $\phi(x)\rightarrow Z^{1/2}\phi_{in}$.

The novel features of {\it Quantum Gravity} call attention on the definition of boundary states and detectors from a different point of view. In general, background independence conflicts with the localization of detectors \cite{m06}. The 2-point function $W[\phi,\Sigma]$ is independent of $\Sigma$, the boundary surface of the interaction region, because it is determined by the gravitational field itself: the distinction between localization and dynamical measurements vanishes.

\section{The Stueckelberg field in the Standard Model}

The Stueckelberg field \cite{Stu,Stu2} is the only up-to-now renormalizable description of a massive vector field (for a review see \cite{sturev,noi}). This formulation is based on describing the massive field by a vector $V_\mu$ and a scalar $\sigma$, so that the full Lagrangian density is given by 
\begin{equation}
\mathcal{L}_{Stueck}=\mathcal{L}_{Proca}(V^\mu)-(\partial^\mu\sigma+mV^{\mu})(\partial_\mu \sigma+mV_\mu)
-(\partial_\mu V^{\mu}+m\sigma)(\partial_\nu V^\nu+m\sigma),
\end{equation}
$\mathcal{L}_{Proca}(V^\mu)$ being the Proca Lagrangian density for $V_\mu$, and $m$ is the mass.

The Stueckelberg field can be introduced in the Standard Model in two different ways:

\begin{enumerate}

{\item as the vector boson $B_\mu$ associated with the hypercharge symmetry $U(1)_Y$,}

{\item as the vector boson associated with an additional Abelian group $U(1)_X$.}

\end{enumerate}

\subsection{$B_\mu$ as the Stueckelberg field}

The Stueckelberg mechanism can apply to the gauge boson associated with the hypercharge symmetry, such that a mass term is naturally predicted for it. Nevertheless, since no extension of this procedure is renormalizable for non-Abelian gauge bosons \cite{DTT}, one still has to refer to the Higgs mechanism to give masses to the $W^i_\mu$ bosons of the $SU(2)$ sector.
 
Hence, in these models \cite{B}, the Stueckelberg mass term adds to that due to the Higgs mechanism and, after the spontaneous symmetry breaking, the Lagrangian density for bosons reads
\begin{equation}
\mathcal{L}_{mass}=\frac{1}{2}\bigg[\frac{g^2v^2}{4}\sum_iW^i_\mu W^{i\mu}-\frac{gg'v^2}{2}W^3_\mu B^\mu+\\+\frac{{g'}^2v^2}{4}B_\mu B^\mu+m^2B_\mu B^\mu\bigg].
\end{equation}

If mass eigenstates $A_\mu$ and $Z_\mu$ are introduced, their masses turn out to be given by the following expressions
\begin{eqnarray}
m^2_Z=\frac{v^2}{8}[(g^2+{g'}^2+\epsilon^2)+((g^2+{g'}^2+\epsilon^2)^2-4g^2\epsilon^2)^{1/2}]\\
m^2_{ph}=\frac{v^2}{8}[(g^2+{g'}^2+\epsilon^2)-((g^2+{g'}^2+\epsilon^2)^2-4g^2\epsilon^2)^{1/2}],
\end{eqnarray}
with $\epsilon=4m^2/v^2$, which show that a massive photon is predicted. 
 
As far as the Weinberg angle is concerned, its expression is given by
\begin{equation}
\tan2\theta=\frac{2gg'}{g^2-{g'}^2-\epsilon^2}
\end{equation}
and this modification reflects in the coupling with fermions, so that a parity-violating interaction between fermions and the photon is predicted. 

All modifications of cross sections with respect to Standard Model can be fixed below the experimental uncertainty, being the photon mass an additional parameter. However, the main criticism to this approach is that it introduces an additional fine-tuned scale.

\subsection{The Stueckelberg field as an additional gauge boson}

The addition of a further $U(1)$ group is the only way to extend the gauge group of the Standard Model, such that anomaly cancellation still occurs. Hence one can develop a gauge theory for the group $SU(3)\otimes SU(2)\otimes U(1)_Y\otimes U(1)_X$.
The Stueckelberg mechanism can apply to the field $C_\mu$ associated with additional sector.

In this respect, the standard assumption \cite{X} is that ordinary matter is neutral with respect to $U(1)_X$, while $C_\mu$ couples with the hypercharge gauge boson $B_\mu$ and with an hidden matter current $J^\mu_X$, according with the following Lagrangian density
\begin{equation}\nonumber
\mathcal{L}=\mathcal{L}_{SM}+\mathcal{L}_{Proca}(C^\mu)-(\partial_\mu\sigma+M_1C_\mu+M_2B_\mu)(\partial^\mu\sigma+M_1C^\mu+M_2B^\mu)+g_XC_\mu J^\mu_X.
\end{equation}

After electro-weak symmetry breaking, one finds
\begin{align}
\mathcal{L}_{mass}&=\frac{1}{2}\bigg[\frac{g^2v^2}{4}\sum_iW^i_\mu W^{i\mu}-\frac{gg'v^2}{2}W^3_\mu B^\mu+\nonumber\\
&+\bigg(\frac{{g'}^2v^2}{4}+M_2^2\bigg)B_\mu B^\mu+M_1^2C_\mu C^\mu+2M_1M_2C_\mu B^\mu\bigg].
\end{align}
 
One can define a rotation in the 3-dim space $(C_\mu,B_\mu,W^3_\mu)$, which diagonalizes the mass matrix. It is worth noting that a massless photon $A_\mu$ still arises, and that it couples with the hidden sector, too, while two massive bosons $Z$ and $Z'$ are predicted.

In this class of models, one deals with two additional parameters $M_1$ and $M_2$, so that it is not difficult to arrange them in order to have agreement with LEP experimental data.
 
In view of an experimental confirmation, the cleanest way to detect the additional sector \cite{X1} would be the emergence of a narrow resonance corresponding to $Z'$. For instance, for a $Z'$ mass $m_{Z'}\approx 100 GeV \div 1 TeV$, the width of the resonance would be $\leq 100 MeV$.
 
There are also attempts to interpret the hidden matter sector as Dark Matter, since these particles turn out to be milli-charged.\\
A recent extension \cite{X2} of this approach is provided by the introduction of the following kinetic mixing terms between $B_\mu$ and $C_\mu$, which arise in some string-like scenarios with orientifolds,
\begin{equation}
\Delta\mathcal{L}=-\frac{\delta}{2}C_{\mu\nu}B^{\mu\nu}.
\end{equation}

It has been shown that, in absence of matter in the hidden sector, modifications of the Standard-Model parameters depend on a single parameter $\bar{\epsilon}=\frac{\epsilon-\delta}{\sqrt{1-\delta^2}}$.

As far as the hidden matter sector is concerned, the presence of the kinetic mixing increases the region of parameters where the relic density is consistent with WMAP data on Dark Matter.

\subsection{Discussion}
We have seen that the Stueckelberg formulation for massive vector fields is currently under investigation, in order to extend the Standard Model of particle physics. 

The application of this mechanism to the hypercharge $U(1)$ gauge group is mainly characterized by a massive photon, which interacts with matter by some parity-violating terms. However, the energy scales at which new phenomena occur can be fixed arbitrary small.

The phenomenology of a Stueckelberg gauge boson associated to an additional $U(1)$ group is richer and more parameters appear. However, in absence of matter from the additional sector, deviations with respect to Standard Model cross section are controlled by a single quantity $\bar{\epsilon}$. Furthermore, as soon as the hidden matter sector is concerned, a new scenario concerning the interpretation of Dark Energy opens.

\section*{Acknowledgments}
We would like to thank G. Montani and R. Ruffini for having attracted our attention on relevant features of Stueckelberg's work.



\begin{thebibliography}{0}    

\bibitem{t46} S. Tomonaga, {\it Prog. Theor. Phys.} 1, 27, (1946); J. Schwinger, {\it Phys. Rev.} {\bf 74} 1439 (1948); F.J. Dyson, {\it Phys. Rev.} {\bf 75} 486 {1949}.
\bibitem{st1} E.C.G. Stueckeberg, T.A. Green, {\it Helv. Phys.Acta} {\bf 24} 153 (1951).
\bibitem{st2} E.C.G. Stueckeberg, A. Petermann, {\it Helv. Phys.Acta} {\bf 26} 499 (1953).

\bibitem{st3} E.C.G. Stueckeberg, {\it Phys. Rev.} {\bf 81} 130 (1951).

\bibitem{d51} F.J. Dyson, {\it Phys. Rev.} {\bf 83} 608 (1951).

\bibitem{mah} G. Kaellen, {\it Helv. Phys. Acta} {\bf 25} 417 (1952); H. Lehmann, {\it Nuovo Cim.} {\bf 11} 342 (1954); H. Lehmann, K. Symanzik and W. Zimmermann, {\it Nuovo Cim.} {\bf 1} 205 (1955); 
C. Itzykson, J.B. Zuber, {\it Quantum Field Theory}  (McGraw-Hill, New York, 1980).

\bibitem{m06} F. Mattei, C. Rovelli , S. Speziale, M. Testa, {\it Nucl.Phys.} {\bf B739} 234 (2006).

\bibitem{Stu}
E.C.G. Stueckeberg, {\it Helv. Phys. Acta} {\bf 11} 299 (1938).

\bibitem{Stu2}
E.C.G. Stueckeberg, {\it Helv. Phys. Acta} {\bf 11} 312 (1938).
 
\bibitem{LB} 
J. H. Lowenstein, B. Schroer, {\it Phys. Rev. } {\bf D6} 1553 (1972). 
 
\bibitem{sturev}
H. Ruegg, M. Ruiz-Altaba, {\it Int. J. Mod. Phys. } {\bf A19} 3265 (2004).

\bibitem{noi} 
F. Cianfrani, O. M. Lecian, {\it Nuovo Cimento B} (2007).

\bibitem{DTT}
R. Delbourgo, G. Thompson, S. Twisk, {\it Int. J. Mod. Phys } {\bf A3} 435 (1988).

\bibitem{B}
S. V. Kuzmin, D. G. C. McKeon, {\it Mod. Phys. Lett. } {\bf A16} 747 (2001).

\bibitem{X}
B. Kors, P. Nath, {\it Phys. Lett. B} {\bf 586} 366 (2004).

\bibitem{X1}
D. Feldman, Z. Liu, P. Nath, {\it JHEP} {\bf 0611} 007 (2006).

\bibitem{X2}
D. Feldman, Z. Liu, P. Nath, {\it Phys. Rev. D} {\bf 75} 115001 (2007).

\end{thebibliography}
\end{document}